\documentclass[12pt]{article}

\usepackage[margin=1.25in]{geometry}
\usepackage{microtype}
\usepackage{fouriernc}
\usepackage{amsmath,amssymb}
\usepackage[svgnames]{xcolor}
\usepackage{hyperref}
\usepackage{orcidlink}
\hypersetup{
    colorlinks=true,
    citecolor=Green,
    urlcolor=Crimson
    }

\title{Algebraic structure behind Odrzywołek's EML operator}
\author{Tomasz Stachowiak\orcidlink{0000-0001-9851-9131}\\
{\small\texttt{tomasz@monodromy.group} }}
\date{}

\begin{document}

\maketitle

\begin{abstract}
\noindent The binary EML operator yields all (transcendental) elementary functions by recursive application, or a binary tree. The structure of the operator itself carries two distinct ingredients: that of an abelian group, and of functional inverse, which reveal a constructive path to many distinct functional families.
\end{abstract}

\section{}

In a recent preprint \cite{Odrz}, Andrzej Odrzywołek described a method of generating all elementary transcendental functions through the single operator
\begin{equation}
    \text{EML}(x,y) = \exp(x) - \ln(y)
\end{equation}
and the constant 1. The construction is recursive and consists in finding deeper and deeper binary trees, whose nodes are copies of this single operator. The author also describes a method of reconstructing specific functions, by employing transcendental constants to verify independence. The set of available functions grows by more or less direct search.

This method is in keeping with the main advantage of EML: that it is a single generative operator, so that a neural network realized as a binary tree with only EML at its nodes can perform symbolic fitting.
However, its brute force nature hides some of the algebraic structure that can be made useful right away: both to extract families of functions for EML, and to give an abstract form to its generalizations.

The motivation for the present note is the fact that the recovery process relies heavily on the ``reductive'' properties of subtraction, and on the addition formula for the exponential.

Crucially, subtraction satisfies: $(y-x)-y = -x$ or, more generally, $x-(y-z) = z-(y-x)$, and this allows us to obtain subtraction and addition, through cancellation.  It is also an essential step that zero is the neutral element, and that the initial constant 1 is chosen so that $\ln(1)=0$. Once 0 becomes available, one can extract each of the component functions: $\exp$ and $\ln$. Together with $\exp(x+y)=\exp(x)\exp(y)$ this is all we need to get multiplication and similarly division.

Let us formalize those intuitive notions somewhat, with a slight abuse of notation: the sets to which the arguments belong are left unspecified, so that the functional composition is purely formal. 
The above description suggests, that the general operator (S for Single) to be scrutinized is of the form
\begin{equation}
    S(x,y) = M\left( f(x), f^{-1}(y)\right),
    \label{gen}
\end{equation}
where $M$ is some operation akin to subtraction or division (EDL). It should thus satisfy three axioms (for all $x,y,z$):
\begin{equation}
\begin{aligned}
    \exists_e\, M(x,e) &= x, &&\text{neutral element}\\
    M(x,x) &= e, &&\text{self-cancellation}\\
    M(x,M(y,z)) &= M(z,M(y,x)), && \text{anti-associativity}
\end{aligned}
\label{axioms}
\end{equation}
where $e$ is the prototype neutral element.

This is not a group operation, but it turns out that there is always a corresponding abelian group behind it. Namely, assuming the axioms, define the binary operation and inverse as
\begin{equation}
\begin{aligned}
    A \boxplus B &:= M(A,M(e,B)),\\
    \iota(A) &:= M(e,A),
\end{aligned}
\end{equation}
and we have an abelian group. In the other direction, let $\boxplus$ be the operation on an abelian group, and define $M$ via
\begin{equation}
    M(A,B) = A \boxplus \iota(B),\\
\end{equation}
to satisfy the axioms \eqref{axioms}, with the shared identity element.

Many suitable binary operations are immediately generated by the conjugation of the ordinary subtraction
\begin{equation}
    M(u,w) = \varphi^{-1}\left(\varphi(u) - \varphi(w)\right).
\end{equation}
Division is the usual example for $\varphi(u)=\ln(u)$; another is $\sqrt{x^2-y^2}$, for $\varphi(u) = u^2$, and it already foreshadows domain problems, which ultimately lead to branch-cuts on $\mathbb{C}$ and manual corrections as noted in \cite{Odrz}. Ultimately, a mild group like $(\mathbb{C},+)$ is broken when mixed with $f=\exp$, which is periodic, so the inverse cancellation in $\ln(\exp(z))=z$ is only formal.

\section{}
Given an appropriate $M$, and the generalized EML of the form \eqref{gen}, we can proceed to function generation.

For brevity, call the inverse function $g$, so that $f(g(x)) = g(f(x)) = x$,
and write the $M$ function as the infix operator $x\boxminus y = M(x,y)$.

The first constant, $c$, has to be given (or otherwise obtained), and it must satisfy 
\begin{equation}
    g(c) = e,
\end{equation}
and the fundamental construction steps of subsequent functions is then the following.
\begin{enumerate}
\item $f_1(x) = f(x)$ by: 
    \[S(x,c) = f(x) \boxminus g(c) = f(x)\boxminus e =f(x).\]
    
\item $f_2(x,y) = f(x) \boxminus y $ by:
    \[S(x,f(y)) = f(x) \boxminus g(f(y)) = f(x)\boxminus y.\]
    
\item $f_3(x) = g(x)$ by:
    \[f_2(z,S(z,x)) = f(z) \boxminus (f(z)\boxminus g(x)) = 
    g(x)\boxminus e = g(x).\]
    
\item $f_4(x,y) = x\boxminus y$ by:
    \[ S(g(x),f(y)) = f(g(x)) \boxminus g(f(y)) = x\boxminus y\]
\item $f_5(x) = \iota(x)$ by:
    \[(x\boxminus y)\boxminus x = 
    (x\boxminus y)\boxminus(x\boxminus e) =
    e\boxminus(x\boxminus(x\boxminus y)) = e \boxminus(y\boxminus e)
    =e \boxminus y\]
    
\item $f_6(x,y) = x\boxplus y$ by:
    \[x\boxminus \iota(y) = x\boxplus y.\]
\end{enumerate}
    
Finally, with a suitable choice of $f$ or $g$ that satisfy an addition formula of the form $f(x\boxplus y) = F(f(x),f(y))$, we obtain further operations by
\begin{equation}
   f(g(x)\boxplus g(y)) = F(f(g(x)), f(g(y)) = F(x,y).
\end{equation}
Most notably, the exponential converts addition to multiplication precisely because $\exp(\ln(x) + \ln(y)) = x y$. It thus plays a double role in the original $\text{EML}$: in itself it generates all the trigonometric functions, but it also translates the additive structure into a multiplicative one. Simple powers, just follow from $\exp(y\ln(x))$.

The particular length of the binary tree for $\ln(x)$ is thus not an inherent feature of the logarithm -- \emph{whatever} the choice of $f$ and $\boxminus$, the recovery of $g$ goes through the third step:
\begin{equation}
    g(x) = S(z, S((S(z,x))),c),
\end{equation}
which, in Polish notation, has length 7.

Note also, that the recovery of $f_5(x)=-x$ could be shorter, by just taking $M(e,x) = S(g(e),f(x))$, but that would be realized through $g^2(c)$, which is $\ln(0)$ for EML. This is not a problem insofar as the implementation admits extended arithmetic with $\infty$, but the derivation above avoids this apparent singularity altogether. 

The concise beauty of the $\text{EML}$ operator is thus that it mixes the core exponential with the logarithm through the anti-associative subtraction. All three can be extracted layer by layer, and then used to reconstruct rational, trigonometric, and power functions.

Yet, the special role of the identity element in how the derivation chain starts, suggests it can't easily be eliminated -- at least not for the general form \eqref{gen}. The only non-generic parameter choice is $S(x,x) = f(x)\boxminus g(x)$, suggesting $f=g$, and thus $f^2=\text{Id}$, which is very restrictive (but see examples below). With ternary operator, on the other hand, it is very easy, as noted in \cite{Odrz}, to immediately combine two operators, e.g.:
\begin{equation}
    B(x,y,z) := \frac{y-z}{x-z}S_1(x,z) + \frac{x-y}{x-z}S_2(x,z),
    \label{twobin}
\end{equation}
which reduces to $B(x,x,z) = S_1(x,z)$ and $B(x,z,z)=S_2(x,z)$. Take $S_1=c$, and we are done. Alternatively, we can accept the constant (it is computationally cheaper after all), and extend the function family with two distinct binary operators instead.

\section{}

There are other, simple starting choices for $f$ and $M$, like the EDL in \cite{Odrz}, which has division as $\boxminus$ and $e=1$. Another uses $g=\exp$ and $c=-\infty$ or $\ln(0)$. They all rely on $\exp$ to recreate the elementary functions, but other choices of $f$ still generate interesting families, perhaps practical in narrower domains as DSL generators.

One such choice is
\begin{equation}
    S(x,y) = \cos(x) - \arccos(y),\quad c=1.
\end{equation}

The fundamental sequence gives not only $\cos(x)$, $\arccos(x)$, and $x\pm y$, but also $\arccos(0)=\pi/2$, from which
$\cos(\tfrac{\pi}{2}-x) = \sin(x)$ follows. Trouble starts with multiplication, though, because the best we can do is
\begin{equation}
    \cos(a+b) + \cos(a-b) = 2\cos(a)\cos(b),
\end{equation}
and making use of the inverse, we get a modified multiplication
\[ F(x,y) = 2xy,\]
but no division, to remove the factor of 2. Ordinary multiplication would require $F(x/2,y)$, i.e. we need other external functions or constants for further extensions. 
Incidentally, we also get the Chebyshev polynomials due to
\begin{equation}
    \cos(n z) = T_n(\cos(z)),
\end{equation}
because multiplication by an integer is realized through addition.

A simple fix is available here by taking $f(x) = 2\cos(x)$, because we then have $f(x+y)+f(x-y) = f(x)f(y)$. Still, despite the functional dependence of $\ln(x)$ on inverse functions like $\arcsin(x)$, we cannot get division, without first having access to roots and rational functions.

Another starting point could be
\begin{equation}
    S(x,y) = \text{arccot}(x) - \cot(y), \quad c=\frac{\pi}{2}.
\end{equation}
Here, we get both $\cot(x)$, $\tan(x)$, $\pm$ (as above) and, because of their relation, also $1/x = \cot(\arctan(x))$. But the addition formula leads even further afield, although it is still a group composition:
\begin{equation}
    \tan(a+b) = \frac{\tan(a)+\tan{b}}{1-\tan(a)\tan(b)}
    \quad\Rightarrow\quad
    F(x,y) = \frac{x+y}{1-xy}.
    \label{cot_law}
\end{equation}
Because we have no access to an operation like division we can't extricate multiplication either. 

As an offshoot, \eqref{cot_law} could be turned into the relativistic velocity addition law if hyperbolic cotangent is used. Alternatively, it is obtained with the simpler pair: $S(x,y) = \tanh(x) - \text{artanh}(y)$ and $c=0$, which generates a ``language'' of the Lorentz boosts (with no constants other than 0).

Next, to link to the problem of a single universal operator, consider the arithmetic on an elliptic curve, which does not allow for rational parametrization, nor for single-function addition formula.

Take the Weierstrass function as the basis so that
\begin{equation}
    S(z,t) = \wp(z) - \wp^{-1}(t), \quad c=\infty.
\end{equation}
Forgetting that the function is doubly periodic and not invertible on its fundamental cell, we can restrict it to some subdomain, and repeat the derivation steps, to recover $\wp(z)$ itself, addition and subtraction.
But even then we hit another obstacle, because the addition formula is of the form
\begin{equation}
    \wp(u+v) = Q(\wp(u),\wp(v),\wp'(u),\wp'(v)), 
\end{equation}
with rational $Q$, so we don't even get a rational composition without the derivative. And it can't be obtained like sine from cosine, through argument shift, because the elliptic curve $y^2=4x^3-g_2x-g_3$ is of genus one, and can't be parametrized rationally or through a single function.

For symmetry's sake we can include 
\begin{equation}
    R(z,t) = \wp'(z) - (\wp')^{-1}(t),
\end{equation}
which makes the rational addition formula for $\wp'$ available. Thus, with two binary operators, we recover the abelian arithmetic on the elliptic curve: $P_1\boxplus P_2=P_3$, where
\begin{equation}
    P_1 = \begin{bmatrix}\wp(u)\\ \wp'(u)\end{bmatrix},\;\;
    P_2 = \begin{bmatrix}\wp(v)\\ \wp'(v)\end{bmatrix},\;\;
    P_3 = \begin{bmatrix}\wp(u+v)\\ \wp'(u+v)\end{bmatrix},
\end{equation}
so that the coordinates of $P_3$ are given through rational functions of the coordinates of $P_1$ and $P_2$.

Finally, to demonstrate a case without external constants, consider the aforementioned possibility of an involutive $f$, for which $S(x,y) = f(x)\boxminus f(y)$ and $S(x,x)=e$. A non-trivial example of a function that is its own inverse, so that $f^2=\text{Id}$, is
\begin{equation}
    f(x) = \left\{ \begin{aligned}
    &\exp(-x)-1, &&\text{for}\; x \geq 0,\\
    &-\ln(x+1), &&\text{for}\; -1 < x < 0.
    \end{aligned} \right.
\end{equation}
The derivation chain for $f(x)-f(y)$ produces $f(x)$ and $\pm$, but the addition formula has a twist. Taking $x=-\ln(t+1)$, and $y=-\ln(s+1)$ to be positive, we have
\begin{equation}
    f(x+y) = \mathrm{e}^{-x}\mathrm{e}^{-y}-1 = (s+1)(t+1) - 1 = ts+t+s,
\end{equation}
Because subtraction is already available, we get multiplication $ts = f(f(s)+f(t))-t-s$, but only for numbers $t$, $s$ between $-1$ and $0$. Not only that, but the only recovered number is $e=c=0$, from which nothing more can be produced via $f$, $\pm$ or multiplication. And this prevents division, because the best we can get is
\mbox{$f(f(s+t)-f(t)) = s/(t+1)$}, with no obvious means of producing 1 to reduce the denominator.

Both this and the ternary $\eqref{twobin}$ smuggle in the IF-THEN-ELSE construction running against the simplicity of EML itself, so the question of a constant-free generator remains open.

\end{document}